# Experimental Study of Coherent Synchrotron Radiation in the Emittance Exchange Line at the A0-Photoinjector

Jayakar C.T. Thangaraj[a], R. Thurman-Keup[a], A. Johnson[a], A. H. Lumpkin[a], H. Edwards[a], J. Ruan[a], J. Santucci[a], Y.E- Sun[a], M. Church[a] and P. Piot[a, b]

[a] *Fermi National Accelerator Labaratory, Batavia, IL,60510, U. S. A*
[b] *Also at the Department of Physics, DeKalb, IL, 60115, U. S. A*

**Abstract.** Next generation accelerators will require a high current, low emittance beam with a low energy spread. Such accelerators will employ advanced beam conditioning systems such as emittance exchanger to manipulate high brightness beams. One of the goals of the Fermilab A0 photoinjector is to investigate the transverse to longitudinal emittance exchange principle. Coherent synchrotron radiation could limit high current operation of the emittance exchanger. In this paper, we report on the preliminary experimental and simulation study of the coherent synchroton radiation (CSR) in the emittance exchange line at A0 photoinjector.

## INTRODUCTION

Modern accelerator based light sources use short electron bunches with high charge and low emittance. To generate short bunches, the electron beam is compressed using one or more bunch compressors. When an electron bunch goes through a bunch compressor, it generates synchrotron radiation in the dipoles. When the wavelength of the synchrotron radiation is larger than the bunchlength, then the radiation becomes coherent and is called as coherent synchrotron radiation (CSR). The intensity of the coherent radiation scales as $N^2$, where N is the number of particles in the bunch. Also, the bunch loses energy due to CSR leading to an energy spread. The power loss due to CSR can be expressed as: $P_{csr} = N^2 x \frac{e^2}{\varepsilon_0 \rho^{2/3} \sigma_z^{4/3}}$, where N is the number of particles, $x$ is 0.0279, $\rho$ is the bending radius and $\sigma_z$ is the bunch length and e is the charge of electron. Therefore, the shorter the bunch at the dipole, the larger will be the power loss due to CSR.

There are other effects as well: The radiation from the tail of the bunch can "catch up" with the head of the bunch and can interact leading to an energy spread within the bunch. The energy spread is then converted to an emittance growth when the bunch exits the bend. This process diminishes the brightness of the beam. Recently, it was found that CSR can also be generated for wavelengths much shorter than the bunch length. If the bunch has micron-scale density modulations due to shot noise, then these modulations generate coherent radiation in the optical and near-IR when the bunch goes through the bend[1]. This can lead to energy modulation that can affect the free electron laser performance. Hence, it is important to investigate the effects of coherent synchrotron radiation effects on a high brightness bunch.

Phase space manipulation of high brightness beam will play a crucial role in future accelerators. One such phase space manipulation involves exchanging the longitudinal emittance with the transverse emittance of the beam. The ability to exchange emittance is useful in free electron lasers where the transverse emittance needs to be very small (~1 micron). Typically, an RF gun can generate a bunch with a smaller longitudinal emittance compared to the transverse emittance. Exchanging the transverse and the longitudinal emittance not only gives a beam with a smaller transverse emittance as required for FELs or colliders but also a larger longitudinal emittance help minimize undesirable effects like the microbunching instability[2].

At the Fermilab A0 photoinjector, a proof of principle experiment on the emittance exchange scheme has been demonstrated[3]. The emittance exchange line consists of a $TM_{110}$ cavity sandwiched by doglegs. The scheme is based on a variation with the original scheme proposed by Cornacchia[4], where the cavity was in the middle of a chicane. A crucial advantage of the new scheme is the output emittances are uncoupled after the exchange. Both the schemes do not take into account collective effects such as space charge effects, coherent synchrotron radiation and wakefields. Our goal is to study coherent synchrotron radiation effects in the emittance exchange line.

The goal of this paper is to answer some questions such as: How does CSR affect the emittance exchange? How can we diagnose the effects of CSR on the beam? In this paper, we report on our measurement of CSR power as a function of charge bunchlength. We then show how CSR could be used to measure the bunchlength of the beam using interferometer. Finally we conclude by proposing techniques for measuring the effects of CSR on the beam.

## EXPERIMENTAL SETUP

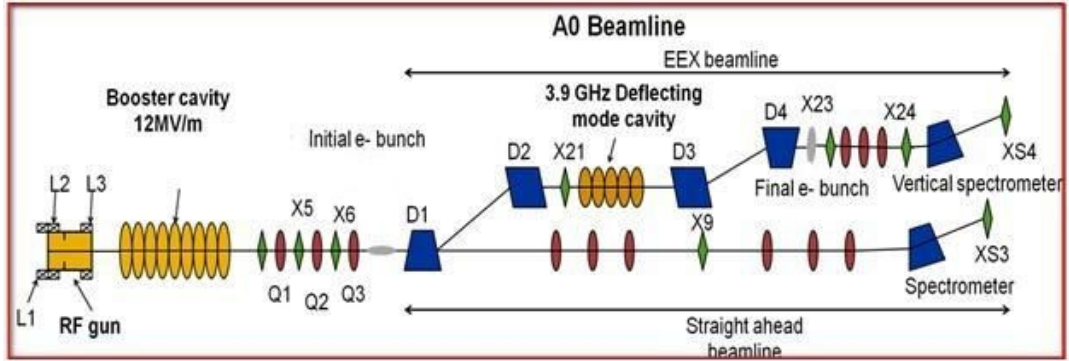

**FIGURE 1.** A schematic of the A0 photoinjector with the emittance exchange line is shown. The 3.9 GHz deflecting mode cavity is switched off in our experiments. The booster cavity phase is adjusted to generate the required chirp on the beam.

Figure 1 shows the emittance exchange facility. The emittance exchange facility consists of the L-band RF gun followed by a superconducting booster cavity, which accelerates the e-beam to 14 MeV. After acceleration, the beam is steered and focused using the dipoles and the quads (Q1 Q2 Q3). The beam then can either continue straight to XS3 or could be bent into the dogleg section of the beamline. In our experiments, the beam is sent through the doglegs (D1 D2 D3 D4) to the spectrometer (XS4). Between the doglegs is the 3.9 GHz deflecting mode cavity, which was switched off in our study.

When the bunch passes through the dogleg, the CSR is expected to be more pronounced at dipole 3. So, we installed optics to collect the radiation coming out of the port at D3. The light is collimated using an off axis parabolic mirror onto a plane mirror. The reflected light is then directed either to a pyrometer (for power measurement) or to an interferometer (to measure bunch length). The interferometer we use is a Martin-Puplett interferometer. The pyrometer we used consists of a single crystalline lithium tantalate[5].

A Martin-Puplett interferometer is a polarizing type interferometer. The incoming light is first sent through a horizontal polarizer. Then, the horizontally polarized light wave is split using a beam splitter and directed to their respective arms of the interferometers. Once split, the light waves travel toward the roof mirror, which changes the polarization of the wave, and reflects them back to the beam splitter, where they are recombined. The recombined wave is again split and sent to pyrodetectors. The pyrodetector signals are recorded on the oscilloscope, which is read by a MATLAB routine. After choosing an appropriate fitting model in the program, the bunch length is calculated. The procedure used to calculate bunchlength and its limitations can be found in [6].

The beam energy for our experiment was around 14MeV and we varied the charge from 250pC to 1 nC. For a given charge, we varied the bunch length by changing the RF off-crest phase in the 9-cell upstream of the dogleg. Since the amount of CSR depends inversely on the bunch length, we used the pyrometer to find a stable operating range of RF off-crest phase.

## RESULTS AND DISCUSSION

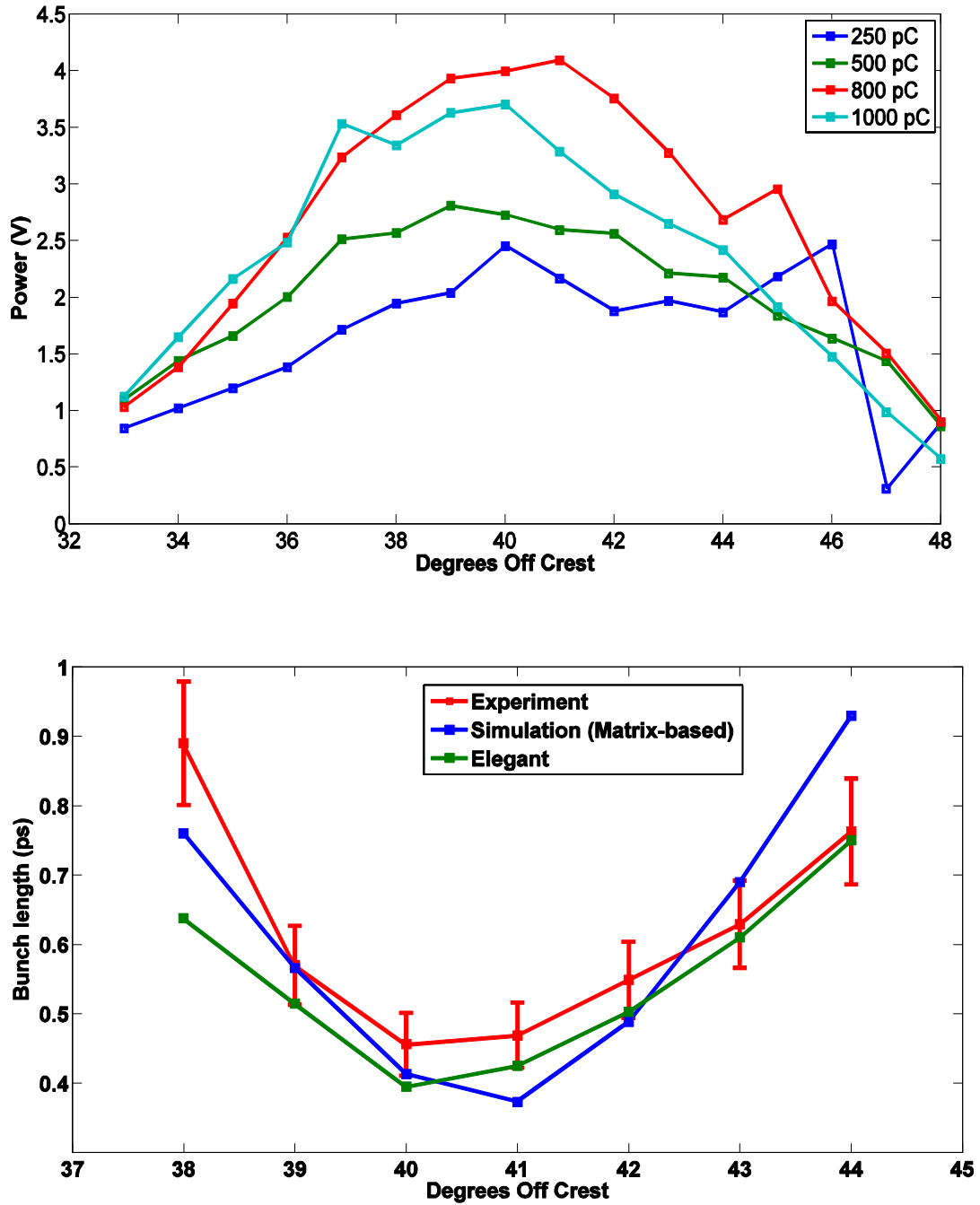

**FIGURE 2.** Experimental results from the 9-cell phase scan. (a) CSR power output measured using pyrometer (b) Bunch length measured using an interferometer showing good agreement with the simulation for shorter bunchlength.

Figure 2(a) shows the plot of the power against the RF off-crest phase for various bunch charges. As we increase the RF phase, the CSR power increases until a peak value after which increasing the RF phase causes a decrease in the CSR power output. The plot also shows that increasing the bunch charge increases the peak value of the CSR power. However, the CSR power for 1nC is less than 800pC. This could be due to the increase in bunchlength because of longitudinal space charge forces. As CSR power is inversely proportional to the bunchlength, increase in bunchlength causes the power to drop. Other reason could be the response of the detector at shorter wavelength.

Here we show why the CSR power peaks at $41^0$ off-crest. The final bunchlength $\sigma_{z,f}$ of the beam going through a bend system can be described as $\sigma_{z,f} = \sqrt{(1 + \kappa R_{56})^2 + (R_{56}\sigma_{z,f}\delta_i)^2}$, where $\sigma_{z,i}$ is the incoming bunchlength. To minimize the bunchlength, we must (a) minimize the incoming energy spread and (b) set $\kappa = -1/R_{56}$. For our bending system, $R_{56} = 0.12$, so the chirp $\kappa = 8.33$. For this chirp, the RF-phase corresponds to $\phi = 23^0$. To minimize the energy spread in our day-to-day operation, we set the RF-phase to $\phi = 23^0$, which corresponds to the chirp of $\kappa = 8.33$. In other words, we should operate the 9-cell phase at a total chirp, $\kappa = 16.6$. This value of chirp is obtained for an RF-phase off crest value of $40^0$, which agrees well with the observed values.

After measuring the CSR power output for various bunch charges, we piped the CSR light into the Martin-Puplett interferometer. We varied the RF-crest phase and recorded the pyrometer signals from the interferometer, which was then used to calculate the bunchlength using MATLAB. This is shown in the figure 2(b). The experimental results are plotted along with the simulation results. Simulations were carried out in *elegant*[7] and in a matrix-based code. There is good agreement between simulation and the experiment for the shorter bunchlength (0.4-0.5 ps) but at longer bunchlength there is a deviation between the simulation and the experiment. One possible reason is that at longer bunchlength it becomes difficult to reconstruct accurately the bunchlength because of the finite detector size. Other reason could be the difference between the real distribution and the assumed beam distribution in the simulation and other initial beam conditions like correlations. The discrepancy between the simulation codes could also be due to the fact that the CSR effects in the dipoles are taken into account in *elegant* while not in the MATLAB code.

## SUMMARY

In this work, we have measured the CSR power as a function of incoming bunchlength. Next, we have shown how CSR could be used to measure the bunchlength, albeit with limitations. In order to study the effects of CSR on the beam, we plan to install a skew quad in the dogleg to measure the longitudinal phase space [8]. Also, we plan to develop single-shot measurement of the bunch length by dispersing the CSR radiation using diffraction gratings and measuring the frequency spectrum.

## ACKNOWLEDGMENTS

We thank the U.S. Department of Energy for supporting our work. The experiment is funded by Fermi Research Alliance, LLC under Contract No. DE-AC02-07CH11359 with the U.S. Department of Energy.

## REFERENCES

1. D. Ratner, A. Chao and Z. Huang, presented at the 30th International Free Electron Laser Conference (FEL 2008), Gyeongju, Korea, 2008.
2. R. Akre, D. Dowell, P. Emma, J. Frisch, S. Gilevich, G. Hays, P. Hering, R. Iverson, C. Limborg-Deprey and H. Loos, Physical Review Special Topics-Accelerators and Beams 11 (3), 30703 (2008).
3. T. Koeth, L. Bellantoni, H. Edwards, R. Filller, A. Lumpkin and J. Ruan, presented at the Particle Accelerator Conference(PAC 09), Vancouver, BC, Canada, 2009 .
4. M. Cornacchia and P. Emma, Physical Review Special Topics - Accelerators and Beams 5 (8), 084001 (2002).
5. InfraTecPyroelectricDetector, http://www.infratec.co.uk/thermography/pyroelectric-detector.html.
6. R. Thurman-Keup, R. P. Fliller and G. Kazakevich, in Beam Instrumentation Workshop (BIW08) ( Lake Tahoe, California, 2008).
7. M. Borland, LS-287, Argonne National Laboratory (2000).
8. K. J. Bertsche, P. Emma and O. Shevchenko, presented at the Particle Accelerator Conference (PAC 09), Vancouver, BC, Canada, 2009.